\documentclass[twoside,11pt]{article}

\usepackage{amstext}
\usepackage{amssymb}

\begin{document}

\title{An Impact of the Metrics Probabilistic Distributions on the Spatial
Geometry of the Universe in Quantum Model}
\author{V.E. Kuzmichev, V.V. Kuzmichev\\[0.5cm]
\itshape Bogolyubov Institute for Theoretical Physics,\\
\itshape Nat. Acad. of Sci. of Ukraine, Metrolohichna Str. 14B, Kiev, 03680 Ukraine}

\maketitle

\begin{abstract}
It is shown that the homogeneous and isotropic Universe is
spatially flat in the limit which takes into account the moments
of infinitely large orders of probabilistic distribution of a
scale factor with respect to its mean value in the state with
large quantum numbers. The quantum mechanism of fine tuning of the
total energy density in the Universe to the critical value at the
early stage of its evolution is proposed and the reason of
possible small difference between these densities during the
subsequent expansion is indicated. A comparison of the predictions
of the quantum model with the real Universe is given.
\end{abstract}

\section{Introduction}

Among the puzzles of the classical cosmology based on the equations of 
general relativity the flatness problem is one of the most important 
\cite{E04,L90,D88}. The model of inflation  \cite{L90, G81} ensures a strict equality
$\Omega = 1$ within the framework of classical cosmology due to
the hypothesis of the De Sitter (exponential) expansion of the early
Universe. The standard $\Lambda$CDM model which includes the inflationary scenario
solves one fine tuning (flatness) problem, but leads to a number of the
new ones (the coincidence between the contributions from dark matter and dark
energy to the total energy density, the smallness of the vacuum energy term
and a requirement for fine tuning of it) \cite{Sh04}.

Measurements of the anisotropy of the cosmic microwave
background (CMB) make it possible to determine the total energy density
$\Omega$ and its components in our Universe. The available astrophysical
data indicate clearly that the modern Universe is very close to be
spatially flat \cite{B00,N02,P02,S03,Sp03,S06}. The results of the WMAP
experiment together with the evidence from the 2dFGRS research and
observations of type Ia supernovae reveal a systematic small deviation
of the total energy density of the Universe in the direction where 
it exceeds a little the critical value \cite{Sp03}. The most accurate
data on the spectra of the CMB fluctuations \cite{H03} were obtained in
the WMAP experiment \cite{S06, Be03, R06}. The position of the first acoustic
peak measured in this experiment, which provides the evidence for
spatial geometry, gives the total energy density equal to $\Omega =
1.003 ^{+ 0.013}_{- 0.017}$ \cite{S06}.

Since a fitting of the values of the free parameters is performed
in multiparametric space, then the possible range of these values
is preassigned in the context of certain assumptions. In this
meaning the interpretation of the WMAP data on the existence and
contributions from separate components in the total energy density
$\Omega$ is model-dependent and may be inadequate to real physical
processes in the expanding Universe. Moreover, secondary effects, 
such as the Sunyaev-Zeldovich effect \cite{SZ69} on the observed CMB anisotropy 
for galaxy clusters at redshift $z \lesssim 1$ \cite{Sh04, M04}
and foreground contamination of the CMB power spectrum from an early epoch
of reionisation at $10 < z < 20$ \cite{Ko03}, might be underestimated.
The search for theoretical models which would provide the higher level of flexibility
with respect to observational cosmology in comparison, e.g., with the standard
$\Lambda$CDM approach is required.

In the present paper a question about the spatial geometry
of the Universe is analyzed on the basis of quantum cosmological
model proposed in \cite{V98,V99a,V99b,VV02}. It has been demonstrated
that the homogeneous and isotropic Universe is spatially flat in 
the limit which takes into account the moments
of infinitely large orders of probabilistic distribution of a
scale factor with respect to its mean value in the state with
large quantum numbers. The quantum mechanism of fine tuning of the
total energy density in the Universe to the critical value at the
early stage of its evolution is discovered and the reason of
possible small difference between these densities during the
subsequent expansion is indicated.

\section{Quantum Model}

As is well known (see, e.g., \cite{M78}), quantum theory 
adequately describes properties of various physical systems.
Its universal validity demands that the Universe as a whole 
must obey quantum laws as well. Since  quantum effects are not 
\textit{a priori} restricted to certain scales \cite{Ki99}, then one should not 
conclude in advance, without research into the properties of the Universe 
within the theory more general than classical cosmology, 
that its space-time structure at large scales will be classical automatically
(the motivation to develop quantum cosmology see in \cite{Ish95,Cou04,VV05}).

The results of the investigations presented in this article are
based on quantum cosmology at the heart of which lies the method
of constraint system quantization proposed by Dirac \cite{D68}
with the addition of the idea of introduction of an additional medium 
or source which determines the reference frame in the Einstein-Hilbert Lagrangian
\cite{V98,VV02,KT91,BK95,BM96}.

As it has been demonstrated in \cite{V98,V99a,V99b,VV02} in quantum theory 
the homogeneous, isotropic and spatially flat Universe filled with the primordial 
matter in the form of a uniform scalar field $\phi$ is described by the
time-dependent Schr\"{o}dinger type equation
\begin{equation}\label{2}
    i\,\partial_{T} \Psi = \hat{\mathcal{H}} \Psi,
\end{equation}
where
\begin{equation}\label{3}
    \hat{\mathcal{H}} = \frac{1}{2} \left(\partial_{a}^{2} -
    \frac{2}{a^{2}}\,\partial_{\phi}^{2} - a^{2} + a^{4} V(\phi)
    \right )
\end{equation}
is a Hamiltonian-like operator, $V(\phi)$ is a potential energy density
of the field $\phi$. Here and below we give all relations between
dimensionless quantities. The length is taken in units of the modified 
Planck length $l_{P} = \sqrt{2 G \hbar/(3 \pi
c^{3})} = 0.744 \times 10^{-33}$ cm, the density is measured in units of
$\rho_{P} = 3 c^{4}/(8 \pi G l_{P}^{2}) = 1.627 \times 10^{117}\,
\mbox{GeV cm}^{-3}$, and so on.

The wavefunction $\Psi$ depends on a cosmological scale
factor $a$, a scalar field $\phi$ and time coordinate $T$
related to the synchronous proper time $t$ by the differential equation
$dt = a\, dT$. When deriving Eq. (\ref{2}) from the principle of least action,
``time'' $T$ is introduced in the theory by means of the coordinate condition
and takes the role of the additional variable which describes the
medium that defines the reference frame \cite{V98,VV02}. In the semi-classical
approach this variable describes the source of the gravitational field
in the form of relativistic matter of an arbitrary nature. Equation (\ref{2})
has a particular solution with separable variables 
\begin{equation}\label{4}
    \Psi = \mbox{e}^{\frac{i}{2} E T} \psi_{E},
\end{equation}
where the function $\psi_{E}$ is defined in the $(a,\phi)$ minisuperspace
and satisfies the time-independent equation
\begin{equation}\label{5}
 \left( -\,\partial _{a}^{2} + \frac{2}{a^{2}}\,\partial _{\phi }^{2} + U - E  \right)
 \psi _{E} = 0,
\end{equation}
while
\begin{equation}\label{6}
U = a^{2} - a^{4} V(\phi)
\end{equation}
can be interpreted as an effective potential. We note that
in the limiting case $E \rightarrow 0$ Eq. (\ref{5}) 
formally turns into the Wheeler-DeWitt equation for the minisuperspace model
\cite{W67}.

Since the Hamiltonian-like operator (\ref{3}) contains
an isotropic oscillator operator with respect to the
variable $a$ as a subsystem, it is convenient to choose
the integration with respect to this variable with a unit
weight function. Using Eq. (\ref{2}) and taking into
account that the operator (\ref{3}) is Hermitian, we obtain
the equation which describes the evolution of the mean value
of some physical quantity represented by the operator $\hat{A}$
in ``time'' $T$,
\begin{equation}\label{7}
    \frac{d}{dT} \langle \hat{A} \rangle = \frac{1}{i}
    \langle [\hat{A},\hat{\mathcal{H}}]\rangle + \langle \partial_{T}\hat{A}
    \rangle,
\end{equation}
where $[\hat{A},\hat{\mathcal{H}}] = \hat{A} \hat{\mathcal{H}} -
\hat{\mathcal{H}} \hat{A}$, and the brackets denote the averaging
over the state $\Psi$ normalized in one way or another (see below). 
Introducing, as usual\cite{LL89}, the operator $d
\hat{A}/dT$, such that
\begin{equation}\label{8}
    \langle \frac{d \hat{A}}{dT}\rangle = \frac{d}{dT} \langle
    \hat{A}\rangle,
\end{equation}
we arrive at the Hiesenberg-type operator equation
\begin{equation}\label{9}
    \frac{d \hat{A}}{dT} = \frac{1}{i}\,
    [\hat{A},\hat{\mathcal{H}}] +
    \partial_{T}\hat{A}.
\end{equation}
Setting $\hat{A} = a$, from Eq. (\ref{9}) we find
\begin{equation}\label{10}
    a\, \frac{d a}{dt} = - \hat{\pi}_{a},
\end{equation}
where $\hat{\pi}_{a} = -i\, \partial_{a}$ is the momentum
operator canonically conjugate with $a$. The operator
equation (\ref{10}) is equivalent to the
definition of the momentum $\pi_{a} = - a\, da/dt$,
canonically conjugate with the variable $a$ in classical cosmology
\cite{L90,V98,VV02}.

Setting $\hat{A} = \hat{\pi}_{a}$, we obtain the equation of 
the evolution of the momentum operator $\hat{\pi}_{a}$ in time $t$
\begin{equation}\label{11}
    a\, \frac{d \hat{\pi}_{a}}{dt} = \frac{2}{a^{3}}\,
    \hat{\pi}_{\phi}^{2} + a - 2 a^{3} V(\phi),
\end{equation}
where $\hat{\pi}_{\phi} = -i\, \partial_{\phi}$ is the momentum
operator canonically conjugate with $\phi$. This equation
is the quantum analog of the canonical equation which determines
the time evolution of the momentum $\pi_{a}$ in classical cosmology.
The momentum of the scalar field, as is well known, equals 
$\pi_{\phi} = \frac{1}{2}\,a^{3}d\phi /dt$. 
The quantum analog of this relation follows from (\ref{9}) at $\hat{A} = \phi$
as well. It has a form 
\begin{equation}\label{12}
    a\, \frac{d \phi}{dt} = \frac{2}{a^{2}}\, \hat{\pi}_{\phi}.
\end{equation}
Using the relations (\ref{10}) -- (\ref{12}), one can obtain
the quantum analogs of all equations of general relativity for
the homogeneous and isotropic Universe filled
with the uniform scalar field and the relativistic
matter.

\section{Choice of Physical States of the Universe}

According to (\ref{5}) the quantum state $\psi_{E}$ depends 
on the form and numerical value of the potential energy density
of the scalar field $V(\phi)$. In the range of values of the
field $\phi$, where the density $V(\phi)$ is the positive-definite
function, the effective potential $U$ (\ref{6}) as a function of  
$a$ at the fixed value of $\phi$ has the form of a barrier.
In this case, the Universe described by Eq. (\ref{5}) can be both in
continuum states with $E > 0$ and quasistationary ones which
correspond to complex values $E = E_{n} + i\,\Gamma_{n}$,
where $E_{n} > 0$, $\Gamma_{n} > 0$ and $\Gamma_{n} \ll E_{n}$,
$n = 0,1,2,\dots$ is the number of a state \cite{V98,V99a,V99b,VV02}.
Quasistationary states are most interesting from the physical
viewpoint, since the Universe in such states can be described 
by a set of standard cosmological parameters accepted in 
classical cosmology (details see in \cite{VV02}). At the same time
the predictions of the quantum model can be compared both with
the predictions of the standard classical cosmology and the data
from astronomical observations.

It can be demonstrated \cite{VV02,V04} that the wavefunction
of a quasistationary state considered as a function of $a$
at the fixed $\phi$ has a sharp peak and is concentrated mainly
in the region limited by the barrier $U$. Then, following Fock
\cite{F76}, one can introduce an approximate function 
$\widetilde{\psi}_{E}$ which is equal to the exact wavefunction 
$\psi_{E}$ inside the barrier and vanishes outside it.
Since the phase of the exact wavefunction $\psi_{E}$ outside
the barrier with respect to $a$ oscillates with the frequency 
that tends to infinity at $a \rightarrow \infty$, and at the same time
its amplitude decreases as $a^{-1}$, in the integrals with
$\psi_{E}$ one can assume that $\psi_{E} \approx
\widetilde{\psi}_{E}$ with a good accuracy. Such an
approximation does not take into account the exponentially
small probability of tunneling through the barrier $U$
in the region of large values of $a$, where $a^{2}\,V > 1$.
It is valid for the calculations of the mean observed parameters 
of the Universe within its lifetime in a given quasistationary state,
when this state can be considered as a stationary one. Here, 
we have a close analogy with the approximate description
of quasistationary states in ordinary quantum mechanics (see,
e.g., \cite{B71}).

In order to determine the character of motion with respect
to the variable $\phi$ we shall use the model of a scalar field which
slowly (in comparison with the rapid, on average, motion with respect to
the variable $a$) rolls from some value $\phi_{start}$ with
the Planck energy density $V(\phi_{start}) \sim 1$ to the 
equilibrium state $\phi_{vac}$ with the energy density
$\rho_{vac} = V(\phi_{vac}) \ll 1$ \footnote{The analogous 
model of a scalar field was considered for the first time
in connection with the inflationary scenario (see, e.g., \cite{L90,LR99}
and references therein).}. This constant density determines
the cosmological constant $\Lambda = 3\,\rho_{vac}$. At the next stage
of the evolution, the scalar field oscillates with a small amplitude
near $\phi_{vac}$ under the action of quantum fluctuations. The
small oscillations of the field $\phi$ near $\phi_{vac}$ can be
quantized \cite{VV04b}. In such a model the motion with respect to $\phi$ 
always will be finite, and the corresponding functions $\psi_{E}$ will
be square-integrable in the $(a,\phi)$ minisuperspace.

\section{Equations for Mean Values}

Performing the averaging over the normalized state (\ref{4}), 
where $\psi_{E} \approx \widetilde{\psi}_{E}$, from Eq. (\ref{5})
we obtain
\begin{equation}\label{13}
    \left \langle \frac{1}{a^{4}}\,\hat{\pi}_{a}^{2} \right
    \rangle = \left \langle \frac{2}{a^{6}}\,\hat{\pi}_{\phi}^{2} \right
    \rangle + \langle V \rangle + \left \langle \frac{E}{a^{4}} \right
    \rangle - \left \langle \frac{1}{a^{2}} \right \rangle.
\end{equation}
In order to reduce this relation to the form which will make it possible
to compare it with the Einstein-Friedmann equation for
the $(^{0}_{0})$ component of classical cosmology, we assume that
in the classical approximation the wave packet represents the
Universe with the scale factor $a_{classic}$ being equal to
the mean value $\langle a \rangle$ in the state $\Psi$, and
the change of position of the packet in time in minisuperspace 
(the expansion or contraction of the Universe in accordance with 
the increasing or decreasing of the scale factor) obeys the laws
of classical cosmology in the limiting case of zero size of the packet.
In agreement with this assumption the Hubble constant will
be determined by the following relation
\begin{equation}\label{14}
    H = \frac{1}{\langle a \rangle}\, \frac{d \langle a
    \rangle}{dt}.
\end{equation}
At such a definition the problems, related with a fact that
the operators $\hat{\pi}_{a}$ and $a$ do not commute between
themselves, do not appear.

Let us extract the contributions from the deviations of $a$
from the mean value $\langle a \rangle$ in an explicit form.
To this end we introduce the operators $\xi$
and $d\xi'/dt$, such that
\begin{equation}\label{15}
    a = \langle a \rangle + \xi,\qquad \frac{da}{dt} =
    \frac{d \langle a \rangle}{dt} + \frac{d\xi'}{dt} .
\end{equation}
Then the relation (\ref{13}) may be reduced to the form
\begin{equation}\label{16}
    H^{2} = \overline{\rho} - \frac{\overline{k}}{\langle a
    \rangle^{2}},
\end{equation}
where we denote
\begin{eqnarray}
  \overline{\rho} &=& \left \langle \left (1 + \frac{\xi}{\langle a \rangle}\right )^{-2}
  \left (1 + \frac{d\xi'}{d\langle a \rangle}\right )^{2}\right \rangle
  ^{-1}\times \nonumber\\
  &\times & \left \{\left \langle \left (1 + \frac{\xi}{\langle a \rangle}\right )^{-6}
  \hat{\pi}_{\phi}^{2}\right \rangle \frac{2}{\langle a \rangle^{6}} \right. + \langle V \rangle + \nonumber\\
    &+& \left.  \left \langle \left (1 +
    \frac{\xi}{\langle a \rangle}\right )^{-4} \right \rangle \frac{E}{\langle a
    \rangle^{4}}\right \},\nonumber \\
    \overline{k} &=& \left \langle \left (1 + \frac{\xi}{\langle a \rangle}\right )^{-2}
  \left (1 + \frac{d\xi'}{d\langle a \rangle}\right )^{2}\right \rangle
  ^{-1}\times \nonumber\\
  \label{17}
  &\times & \left \langle \left (1 + \frac{\xi}{\langle a \rangle}\right
  )^{-2}\right \rangle.
\end{eqnarray}
This equation is an exact expression. It takes into account 
all quantum corrections with respect to the deviation $\xi$.
In zero approximation $\xi = 0$, and the change of the mean
$\langle a \rangle$ in time $t$ is determined by the equation
\begin{equation}\label{18}
    H^{2} = \langle \rho \rangle - \frac{1}{\langle a
    \rangle^{2}},
\end{equation}
where
\begin{equation}\label{19}
    \langle \rho \rangle = \frac{2}{\langle a \rangle^{6}} \langle \pi_{\phi}^{2}
    \rangle + \langle V \rangle + \frac{E}{\langle a \rangle^{4}}.
\end{equation}
This equation may be considered as the Einstein-Friedmann equation
in terms of mean values. The quantity $\langle \rho \rangle$
gives the mean total energy density in the Universe filled
with the scalar field and the relativistic matter.

In accordance with the correspondence principle which establishes
an agreement between the quantum and classical descriptions
of the physical system (see, e.g., \cite{M78}), in Eq. (\ref{18}) 
the mean values should be calculated in the state with large 
quantum numbers. Such a state is described by the wavefunction 
$\psi_{E}$ with separable variables,
\begin{equation}\label{20}
    \psi_{E}(a,\phi) = \varphi_{n}(a)\, f_{ns}(\phi).
\end{equation}
(An explicit form of $\varphi_{n}$ and $f_{ns}$ is given in
\cite{V99b,VV02} for $\phi_{vac} = 0$ and in \cite{VV04} in the
general case.) Here, the quantum number $n$ describes the number
of elementary quantum excitations of the vibrations of oscillator
which characterizes a variation of the metric (their number is
equal to $N = 2n + 1$), and $s$ characterizes the number of the
elementary quantum excitations of vibrations
of the scalar field near the equilibrium state $\phi_{vac}$.
The latter excitations can form an invisible energetic component
in the total energy density in the Universe \cite{VV04b}.

The mean density (\ref{19}) in the state (\ref{20}) equals
\begin{equation}\label{21}
    \langle \rho \rangle = \gamma \frac{M}{\langle a \rangle^{3}}
    + \rho_{vac} + \frac{E}{\langle a \rangle^{4}},
\end{equation}
where $\gamma = 193/12$ is a numerical coefficient which
appears in the calculation of expectation values of the operators
of the kinetic and potential parts of the energy density of
the scalar field in expression (\ref{19}), $M = m\,(s + 1/2)$
can be interpreted as the amount of matter/energy in the Universe
represented in the form of a sum of the elementary quantum excitations of  
vibrations of the field $\phi$ with the masses $m =
([\partial_{\phi}^{2} V]_{\phi_{vac}})^{1/2}$.

\section{Quantum Corrections}

Let us calculate the quantum corrections to the classical
density (\ref{21}), using the exact expression for 
$\overline{\rho}$ from (\ref{17}). We assume that 
$d\xi'/dt \ll d\langle a \rangle/dt$. This corresponds to the case,
when the deviation $\xi$ depends weakly on the mean value 
$\langle a \rangle$ (i.e., the corresponding statistical distribution
slowly changes in the form during the small time intervals).
According to Eq. (\ref{16})
the quantity $\overline{\rho}$ can be considered as the energy density
which takes into account the quantum corrections. For the states (\ref{20})
it can be reduced to the form
\begin{equation}\label{22}
    \overline{\rho} = \frac{1}{Z_{2}}\left\{Z_{6}\, \frac{2}{\langle a \rangle^{6}}\, \langle \pi_{\phi}^{2}
    \rangle + \langle V \rangle + Z_{4}\, \frac{E}{\langle a
    \rangle^{4}}\right\},
\end{equation}
where we denote
\begin{equation}\label{23}
    Z_{l} = \left\langle \left(1 + \frac{\xi}{\langle a \rangle}\right)^{-l}
    \right\rangle.
\end{equation}
The quantities $Z_{l}$ play the role of the ``renormalization constants''.
They may be rewritten in the form of the infinite alternating series
\begin{equation}\label{24}
    Z_{l} = 1 + \sum_{\mu = 2}^{\infty} (-1)^{\mu}\, \frac{l (l + 1) \cdots (l + \mu - 1)}
   {\mu \,!}\, \frac{\langle \xi ^{\mu }\rangle}{\langle a \rangle ^{\mu }}.
\end{equation}
Here, the mean $\langle \xi^{2} \rangle = \langle a^{2} \rangle -
\langle a \rangle^{2}$ is the dispersion, and $\langle \xi^{\mu} \rangle
= \langle(a - \langle a \rangle )^{\mu} \rangle$ at $\mu
> 2$ determines the moment of order $\mu$ of probabilistic distribution of a
scale factor with respect to its mean value $\langle a \rangle$. For the states
with $n \gg 1$ we find
\begin{eqnarray}
    \frac{\langle \xi^{\mu}\rangle}{\langle a \rangle^{\mu}} &=& \frac{1}{\mu +
    1}\quad \mbox{for even numbers of} \ \mu, \nonumber \\
    \label{25}
    \langle \xi^{\mu}\rangle &=& 0 \quad \mbox{for odd numbers of} \ \mu.
\end{eqnarray}
In this case the constants $Z_{l}$ will be given by the asymptotic series
\begin{equation}\label{26}
    Z_{l} = 1 + \sum_{\mu = 2}^{\infty}\! '\, \frac{l (l + 1) \cdots (l + \mu - 1)}{(\mu +
    1)!},
\end{equation}
where the prime near the summation sign means that the summation
is performed only with respect to the even numbers of $\mu$. For the Universe
in the states with $n \gg 1$ and $s \gg 1$ from (\ref{22}) we obtain
\begin{equation}\label{27}
    \overline{\rho} = \left(1 + \frac{\Delta \rho}{\langle \rho
    \rangle}\right) \langle \rho \rangle,
\end{equation}
where the quantum correction
\begin{eqnarray}
    \Delta \rho &=& \left[\left(\frac{Z_{6}}{Z_{2}} - 1\right) 16 +
    \left(\frac{1}{Z_{2}} - 1\right) \frac{1}{12} \right] \frac{M}{\langle a
    \rangle^{3}} + \nonumber \\
    \label{28}
    &+& \left(\frac{1}{Z_{2}} - 1\right) \rho_{vac} +
    \left(\frac{Z_{4}}{Z_{2}} - 1\right) \frac{E}{\langle a \rangle^{4}}
\end{eqnarray}
takes into account the contributions from the dispersion and all nonzero
moments $\langle \xi^{\mu} \rangle$ into the dynamics of the Universe.

In the case, when the contributions from the vacuum and relativistic matter
may be neglected, 
\begin{equation}\label{29}
    \rho_{vac} \sim 0 \quad \mbox{and} \quad \frac{E}{\langle a
    \rangle^{4}} \sim 0,
\end{equation}
the relative correction to the density $\langle \rho \rangle$
is expressed only in terms of the renormalization constants $Z_{l}$,
\begin{equation}\label{30}
    \frac{\Delta \rho}{\langle \rho \rangle} = \frac{1}{\gamma}
    \left[\left(\frac{Z_{6}}{Z_{2}} - 1\right) 16 +
    \left(\frac{1}{Z_{2}} - 1\right) \frac{1}{12} \right].
\end{equation}

In accordance with Eq. (\ref{16}) the density parameter $\Omega$
at $\overline{k} = 1$ is determined by the expression
\begin{equation}\label{31}
    \Omega = \frac{\overline{\rho}}{H^{2}}.
\end{equation}
Then, taking into account (\ref{27}), from (\ref{16}) we obtain
\begin{equation}\label{32}
    \Omega = \left[1 - \frac{1}{\langle a \rangle^{2} \langle \rho
    \rangle}\left(\frac{1}{1 + \frac{\Delta \rho}{\langle \rho
    \rangle}}\right)\right]^{-1}.
\end{equation}
There exists the constraint equation $\langle a \rangle = M$ between
the geometry and matter in the approximation (\ref{29}). This condition
is the particular case of a more general feedback coupling relation 
between the geometric and energetic characteristics of the Universe
\begin{equation}\label{33}
    \langle a \rangle = M + \frac{E}{4 \langle a \rangle} + 4 \langle a
    \rangle^{3}\rho_{vac},
\end{equation}
where the second term on the right-hand side describes
the energy of a relativistic matter, while the third term gives the
contribution from the vacuum of the scalar field. It follows from
the condition on eigenvalues $E$ of Eq. (\ref{5}) for the states 
with $n \gg 1$ and $s \gg 1$,
\begin{equation}\label{34}
    E = 2N - (2N)^{2} \rho_{vac} - 2 \sqrt{2N}\,M,
\end{equation}
where $N = 2n + 1$, and the mean $\langle a \rangle = \sqrt{N/2}$
\cite{VV05}. This equation must be taken into account in the
calculations of the expectation values of observed parameters.

\begin{table}
  \centering
\caption{The deviation of $\Omega$ from unity depending on
 the number of terms which are taken into account in the sum over
 $\mu$ in Eq. (\ref{26}); $\mu_{max}$ is the largest order of
 the moments $\langle \xi^{\mu}\rangle$ taken into account in the correction
 (\ref{28}). The cosmological constant $\Lambda$ was determined according to 
 the type Ia supernovae data \cite{R04}.}\label{01}

\vspace{0.3cm} 
\begin{tabular}{|c|c|c|c|}
  \hline
    & \multicolumn{2}{|c|}{$\Omega - 1$} & \\
  \cline{2-3}
  $\mu_{max}$ & $\Lambda = 0$ & $\Lambda \neq 0$ & $-\,\Lambda \times 10^{58},\, \mbox{cm}^{-2}$\\
  \hline
  0 & $6.63 \times 10^{-2}$ & $3.47 \times 10^{-2}$ & $1.11$               \\
  2 & $1.59 \times 10^{-2}$ & $8.30 \times 10^{-3}$ & $2.78 \times 10^{-1}$\\
  4 & $5.68 \times 10^{-3}$ & $2.95 \times 10^{-3}$ & $1.00 \times 10^{-1}$\\
  6 & $2.53 \times 10^{-3}$ & $1.31 \times 10^{-3}$ & $4.48 \times 10^{-2}$\\
  8 & $1.29 \times 10^{-3}$ & $6.72 \times 10^{-4}$ & $2.29 \times 10^{-2}$\\
 10 & $7.28 \times 10^{-4}$ & $3.79 \times 10^{-4}$ & $1.29 \times 10^{-2}$\\
 12 & $4.42 \times 10^{-4}$ & $2.30 \times 10^{-4}$ & $7.84 \times 10^{-3}$\\
 14 & $2.83 \times 10^{-4}$ & $1.47 \times 10^{-4}$ & $5.03 \times 10^{-3}$\\
 \hline
\end{tabular}

\end{table}

In Table 1 we give the deviation of $\Omega$ from unity for different 
approximations with respect to the constants $Z_{l}$, which take into account
the terms up to the moment of order $\mu_{max}$ in the sum over $\mu$ in 
(\ref{26}). For example, $\mu_{max} = 0$ corresponds to the case
$Z_{l} = 1$ and is described by a zero approximation (\ref{18}), (\ref{21}).
The value $\mu_{max} = 2$ corresponds to the case, when one term (dispersion) 
with $\mu = 2$ is taken into consideration, $\mu_{max} = 4$ accounts for
two terms (dispersion and fourth moment) with $\mu = 2,4$, and so on.
The column with $\Lambda = 0$ corresponds to the condition (\ref{29}), 
the cosmological constant $\Lambda \neq 0$ was determined according to 
the type Ia supernovae data \cite{R04}.
It is interesting to note that for $\rho_{vac} = 0$ taking the dispersion 
into account leads to the value $\Omega = 1.016$ that is in good agreement 
with the WMAP data. The astrophysical data obtained previously, 
$\Omega = 1 \pm 0.12$ \cite{B00}, $\Omega = 1.02
\pm 0.06$ \cite{N02}, $\Omega = 1.04 \pm 0.06$ \cite{P02}, $\Omega
= 0.99 \pm 0.12$ \cite{S03}, are described by a zero approximation.

Let us consider the case, when $\rho_{vac} \neq 0$, but the
contribution from the relativistic matter will be neglected as before.
We determine a single free parameter of the theory $\Lambda$ 
from a $\chi^{2}$ statistic for the distance modulus of
the source as a function of the cosmological redshift $z = a_{0}/ \langle a \rangle - 1$, 
where $a_{0}$ is the scale factor at the moment of observation. We take
156 type Ia ``gold'' supernovae as the sources with the different $z$ \cite{R04}.
The results of such analysis with $\chi_{dof}^{2} = 1.17$ are given in Table 1.
They demonstrate that the cosmological constant $\Lambda$ in this theory 
is negative in all approximations. While one takes into account the contributions 
from quantum corrections of higher and higher orders of $\mu$, it diminishes.
At the same time the scale factor $a_{0}$ grows so that the value $a_{0}^{2}\,\Lambda$ remains
almost constant being close to the limiting value $a_{0}^{2}\,\Lambda = -0.692$ for 
$\mu_{max} \geq 10$. 

We note that the idea of
occupied levels with negative energy \cite{D79} leads to a negative
energy density as well \cite{ZN71}. Moreover, superstring models
of quantum gravity which invoke compactified higher spatial dimensions
are incompatible with the positive cosmological constant of the model with
the cold dark matter and prefer models with negative or no
cosmological constant \cite{Sh04}.

\section{An Asymptotic Limit of the Spatial Geometry}

Since the renormalization constants $Z_{l}$ are described by the
asymptotic series (\ref{26}) which give a finite result in
every approximation, then, generally, in the limit which takes into account
the moments of arbitrarily large but finite orders $\langle \xi^{\mu}
\rangle$, we obtain $\Omega = 1 + \varepsilon$, where $\varepsilon
\sim +0$. In other words, the quantum model predicts an arbitrarily small
but finite excess of the density $\Omega$ over unity in the homogeneous and 
isotropic Universe. This agrees with the basic premise (Eq. (\ref{5})
describes the spatially closed Universe). As was noted in Introduction,
the data of the CMB anisotropy observations most likely point out a
small enough but systematic excess of the current energy density
in the Universe over its critical density \cite{Sp03, S06}.

In the limit $\mu \rightarrow \infty$ (for an infinitely large
number of terms of the asymptotic series (\ref{26})) we obtain
an exact expression, $\Omega = 1$. This means that from the standpoint
of the quantum description the Universe will be spatially flat
in the epoch, when arbitrarily large, on average, deviations of
the scale factor $a$ from the mean value $\langle a \rangle$ are
possible. The assumption that the early Universe must obey the quantum
laws to a greater extent then the classical ones seems justified.
Then from general physical reflections it is clear that in the early epoch,
when nevertheless the state of the Universe may be characterized by
the large quantum number $n$ \footnote{Since the motion with
respect to the variable $a$ is described by an oscillator, then
$\langle a \rangle \sim \sqrt{n}$. From this it follows, in particular,
that in the state with $n \sim 10$ the ``radius'' of the Universe
is still close to the Planck value $\langle a \rangle
\sim 1$.}, such deviations are most probable.

This result agrees completely with the conclusions of general relativity
that the early Universe must be spatially flat to a higher accuracy,
then nowadays \footnote{Estimations within general relativity
give the values $|\Omega - 1| \sim 10^{-60}$ for $t \sim 10^{-44}$ s, $|\Omega -
1| \sim 10^{-20}$ for $t \sim 10^{-10}$ s, and $|\Omega - 1| \sim
10^{-1} - 10^{-2}$ for the current epoch with $t \sim 10^{10}$ years
\cite{L90,D88}.}. Thus the quantum model points out the natural mechanism
of fine-tuning of the parameter $\Omega$ to unity at early stages
of the evolution of the Universe, as general relativity demands, and
the reason for a small possible difference of the 
energy density from the critical value in process of subsequent expansion.

\vspace{0.3cm}
\noindent \textbf{Acknowledgments}\\
This work was supported partially by the Program of Fundamental Research
``The Fundamental Properties of Physical Systems under Extremal Conditions''
of the Physics and Astronomy Division of the National
Academy of Sciences of Ukraine. 

{}

\end{document}